\pdfoutput=1
\documentclass[12pt,preprint]{aastex}
\usepackage{graphicx}
\usepackage{natbib}
\usepackage{aas_macros}
\usepackage[section] {placeins}
\usepackage{subfigure}
\usepackage{float}
\usepackage{color}

\usepackage[scaled]{helvet}

\usepackage[T1]{fontenc}

\def\msun{{\rm\,M_\odot}}

\def\msun{{\rm\,M_\odot}}

\newcommand{\kms}{\, {\rm km\, s}^{-1}}

\newcommand{\mpc}{\, {\rm Mpc}}
\newcommand{\kpc}{\, {\rm kpc}}
\newcommand{\hmpc}{\, h^{-1} \mpc}

\newcommand{\hi}{\mbox{H\,{\scriptsize I}\ }}
\newcommand{\heii}{\mbox{He\,{\scriptsize II}\ }}

\def\h2{${\rm\,H_2}$}

\makeatletter 

\def\hi{\noindent \hangindent=2.5em}

\def\kpc{{\rm\,kpc}}

\def\mpc{{\rm\,Mpc}}

\def\kms{{\rm\,km/s}}
\def\msun{{\rm\,M_\odot}}

\def\vol#1  {{{#1}{\rm,}\ }}

\def\hi{{H\thinspace I}\ }

\def\heii{{He\thinspace II}\ }

\newcount\refno
\newcount\rfno
\def\eq{$^{\the\refno\ }$\advance\refno by 1}
\def\ad{\advance\rfno by 1}

\def\clock{\count0=\time \divide\count0 by 60
     \count1=\count0 \multiply\count1 by -60 \advance\count1 by \time
     \number\count0:\ifnum\count1<10{0\number\count1}\else\number\count1\fi}

\def\myputfigure#1#2#3#4#5%
{\hskip0.03\textwidth\vskip#5pt
\makebox[0pt]{\hskip#2in
\includegraphics[width=#3\textwidth]{#1}}\vskip#4pt\hfill}

\newcount\refno
\newcount\rfno
\def\eq{$^{\the\refno\ }$\advance\refno by 1}
\def\ad{\advance\rfno by 1}

\definecolor{burntorange}{rgb}{1,0.4,0.2}

\definecolor{burntorange}{rgb}{1,0.4,0.2}

\usepackage{color}

\begin{document}

\title{Testing Dark Matter Halo Models of Quasars With Thermal Sunyaev-Zeldovich Effect}

\author{Renyue Cen\altaffilmark{1} and Mohammadtaher Safarzadeh\altaffilmark{2}}

\footnotetext[1]{Princeton University Observatory, Princeton, NJ 08544;
 cen@astro.princeton.edu}

\footnotetext[2]{Johns Hopkins University, Department of Physics and Astronomy, Baltimore, MD 21218, USA}

\begin{abstract} 

A statistical analysis of stacked Compton$-y$ maps of quasar hosts with a median 
redshift of $1.5$ using {\em Millennium Simulation} is performed to address two issues,
one on the feedback energy from quasars and the other on testing dark matter halo models for quasar hosts.
On the first, we find that, at the resolution of FWHM=$10$ arcmin obtained by Planck data,
the observed thermal Sunyaev-Zeldovich (tSZ) effect can be entirely accounted for and explained by 
the thermal energy of halos sourced by gravitational collapse of halos,
without a need to invoke additional, large energy sources, such as quasar or stellar feedback.
Allowing for uncertainties of dust temperature in the calibration of observed Comton$-y$ maps,
the maximum additional feedback energy is $\sim 25\%$ of that previously suggested.
Second, we show that, with FWHM=$1$ arcmin beam, tSZ measurements will provide 
a potentially powerful test of quasar-hosting dark matter halo models,
limited only by possible observational systematic uncertainties, not by statistical ones,
even in the presence of possible quasar feedback. 
\end{abstract}

\section{Introduction}

The nature of the dark matter halos hosting quasars remain debatable.
There are primarily two competing models. One is the traditional, popular
HOD (halo occupation distribution) model, which is based on assigning a probability function to quasars
to reside in a halo of a given mass in order to match the observed quasar clustering strength \citep{2005Zheng,2007Zheng,2013Shen}.
The other model is a physically motivated model recently put forth \citep[]['CS model' hereafter]{2015Cen}.

While the CS model, like the HOD based model, matches the observed clustering of quasars,
the masses of the dark matter halos in the CS model are very different from those of the HOD based model.
For example, at $z\sim 0.5-2$, the host halos in the CS model have masses of $\sim 10^{11}-10^{12}\msun$,
compared to $(0.5-2)\times 10^{13}\msun$ in the HOD model.
This then offers a critical differentiator between the CS and HOD models, namely, 
the cold gas content in quasars host galaxies.
Specifically, because of the large halos mass required in the HOD model,
quasars hosts have much lower content of cold gas than in the CS model.
\citet[][]{2015Cen} have shown that
the CS model is in excellent with the observed 
covering fraction of $60\%-70\%$ for Lyman limit systems within the virial radius of $z\sim 2$ quasars \citep[][]{2013Prochaska}.
On the other hand, the HOD model is
inconsistent with observations of the high covering fraction of Lyman limit systems in quasar host galaxies. 
Given the fundamental importance of the nature of dark matter halos hosting quasars,
in this {\it Letter} we present another potentially powerful test to distinguish between these two competing models.
We show that upcoming measurements of thermal Sunyaev-Zeldovich effect at arc-minute resolution (or better)
should be able to differentiate between them with high confidence.

\section{Simulations and Analysis Method}\label{sec: sims}

We utilize the {\em Millennium Simulation} \citep[][]{2005dSpringel} to perform the analysis. 
A set of properties of this simulation that meet our requirements includes
a large box of $500h^{-1}$Mpc,
a relatively good mass resolution with dark matter particles of mass $8.6 \times 10^8 h^{-1}\msun$,
and a spatial resolution of 5 $h^{-1}\,{\rm kpc}$ comoving.
The mass and spatial resolutions are adequate for capturing halos of masses greater than $10^{11}\msun$,
which are resolved by at least about 100 particles and 40 spatial resolution elements for the virial diameter.
Dark matter haloes are found through a {\em friends-of-friends} (FOF) algorithm.
Satellite halos orbiting within each virialized halo are identified applying a SUBFIND algorithm \citep[][]{2001dSpringel}. 
The adopted $\Lambda{\rm CDM}$ cosmology parameters are
$\Omega_{\rm m}=0.25$, $\Omega_{\rm b}=0.045$, $\Omega_{\Lambda}=0.75$,
$\sigma_8=0.9$ and $n=1$, where the Hubble constant is $H_0=100{h\,\rm km}
\,{\rm s}^{-1}\,{\rm Mpc}^{-1}$ with $h=0.73$. 
We do not expect that our results strongly depend on the choice of cosmological parameters
within reasonable ranges, such as those from \citet{2011Komatsu}.

The steps taken to construct the tSZ maps are as follow.
For each model (either CS or HOD) quasar model, we sample the quasar host dark matter halos at each redshift,
$z=0.5$, $1.4$ and $3.2$.
For each quasar host, we select all halos within a projected radius of 80 arcmin centered at the quasar 
in a cylinder with the depth equal to the length of the simulation box in a given direction. 
The thermal energy of a halo of mass $M_h$ is calculated using
\begin{equation}
\label{eq:E}
{\rm E_{th}= \frac {3 \Omega_{b}}{2 \Omega_{m}} M_h \sigma^2} 
\end{equation}
\noindent
where $M_h$ is the halo mass 
and $\sigma$ the 1-d velocity dispersion computed as 
\begin{equation}
\label{eq:sigma}
{\rm \sigma=0.01\times \Big(\frac{M_h}{M_\odot}\Big)^{1/3} \Big[{\frac{\Omega_{M}(z=0)}{\Omega_M (z)}}\Big]^{1/6} (1+z)^{1/2}} {\rm [km/s]}.
\end{equation}
\noindent
The energy of each halo is then distributed uniformly in projected area inside its virial radius ${\rm r_{v}}$.
To construct SZ maps, we project the energy of each halo using a cloud-in-cell technique in 2-d.
We obtain the Compton-y parameter corresponding to total projected thermal energy ${\rm E_{th}/A}$ at each pixel with:
\begin{equation}
\label{eq:prob}
{\rm y = 0.88\times0.588\times \frac{2 \sigma_T E_{th}}{3 m_e c^2 A} }
\end{equation}
\noindent
where ${\rm A}$ is the area of the pixel, 
${\rm \sigma_T}$ the Thomson scattering cross section, 
${\rm m_e}$ the electron mass, 
${\rm c}$ the speed of light, 
and $0.88$ and $0.58$ accounts for electron density to mass density, and molecular weight, respectively.
We limit the dark matter halos that contribute to the $y$ calculation
to the mass range $[3\times10^{12}, 5.5\times10^{14}]\msun$ at $z=0.5$ 
and  $[3\times10^{12}, 6.5\times10^{14}]\msun$ at both $z=1.4$ and $3.2$. 
The upper mass limits is used in order to enable comparisons to observations,
accounting for the fact that in the Planck observation generated $y$ maps the clusters more massive
than these indicated upper limits are masked out \citep[][]{2014Planck}.
The lower mass limits reflect the fact that less massive halos would be
cold stream dominated instead of virial shock heated gas dominated;
changing the lower mass limit from 
$3\times10^{12}\msun$ to $1\times10^{12}\msun$ only slightly increases the computed y parameter.

To enable comparison with the observed Compton-y maps stacked over a range of redshift $z\sim 0.1-3.0$
with median redshift of $z_{med}\sim 1.5 $ \citep[][]{2015Ruan},
we appropriately assign weightings of (36\%, 51\%, 13\%) for $z=(0.5, 1.4, 3.2)$ maps, respectively,
and sum up the contributions from the three redshifts.
These weightings are adopted to mimic the redshift distribution of stacked quasars used in the observational anaylysis.
To compute the variance of the y-parameter we make nine maps each averaged over 1/9th of total individual Compton-y maps of the quasar hosts 
at each redshift for either of the HOD or CS models. Then we have $9\times9\times9=729$ possible final maps constructed with the weightings defined above.
The dispersion and the mean is then computed considering these 729 final maps. 
In addition, we construct isolated quasar host only $y$ maps, with only the quasar host halo's energy contributing to the final tSZ map. 
In other words, in those isolated quasar y maps, we exclude effects from projected, clustered neighboring halos.

\section{Validating Quasar Models with Planck Thermal Sunyaev-Zeldovich Effect Maps}

\begin{figure}[!h]
\centering
\vskip -0.0cm
\resizebox{3.0in}{!}{\includegraphics[angle=0]{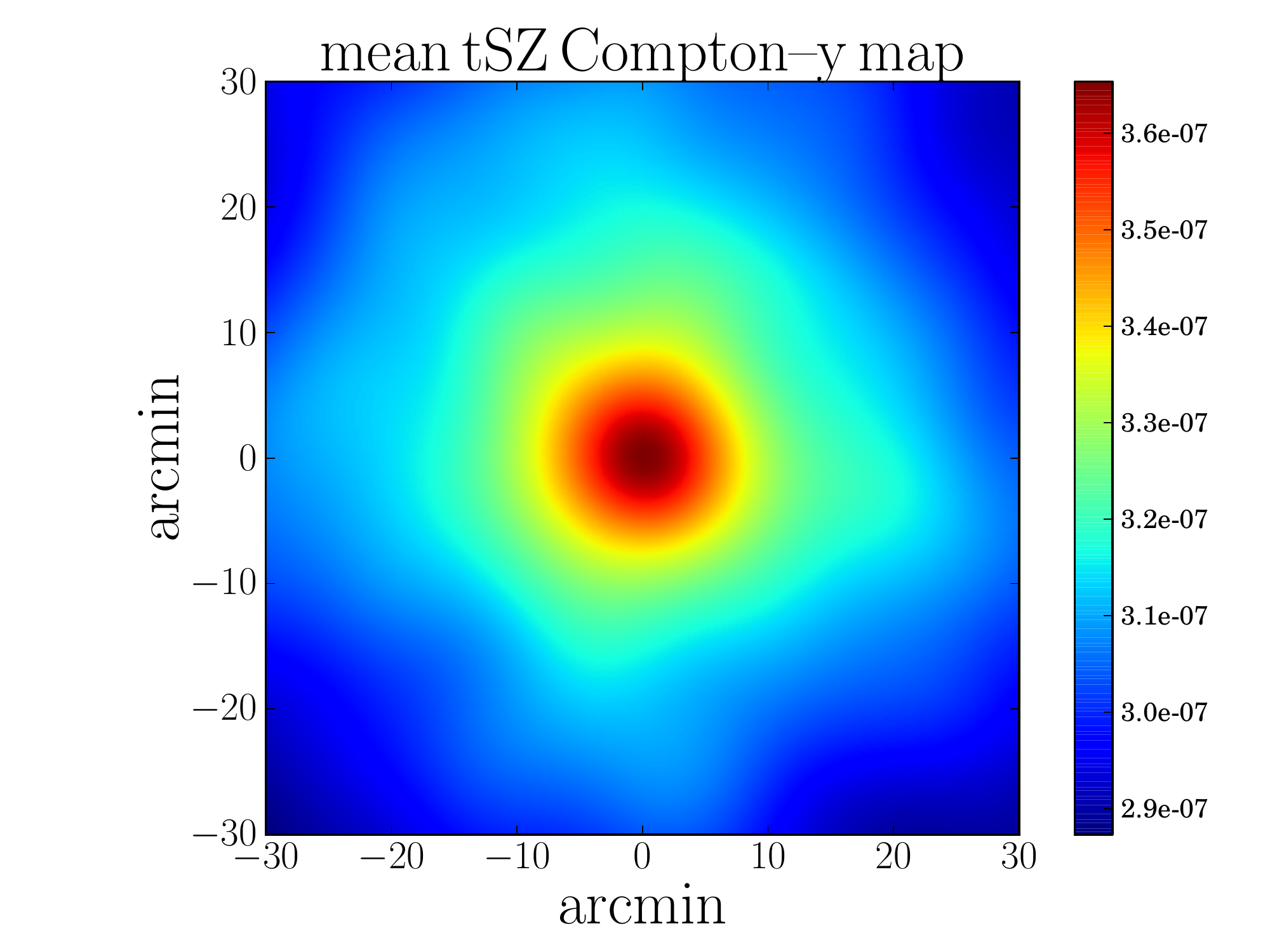}}
\resizebox{3.0in}{!}{\includegraphics[angle=0]{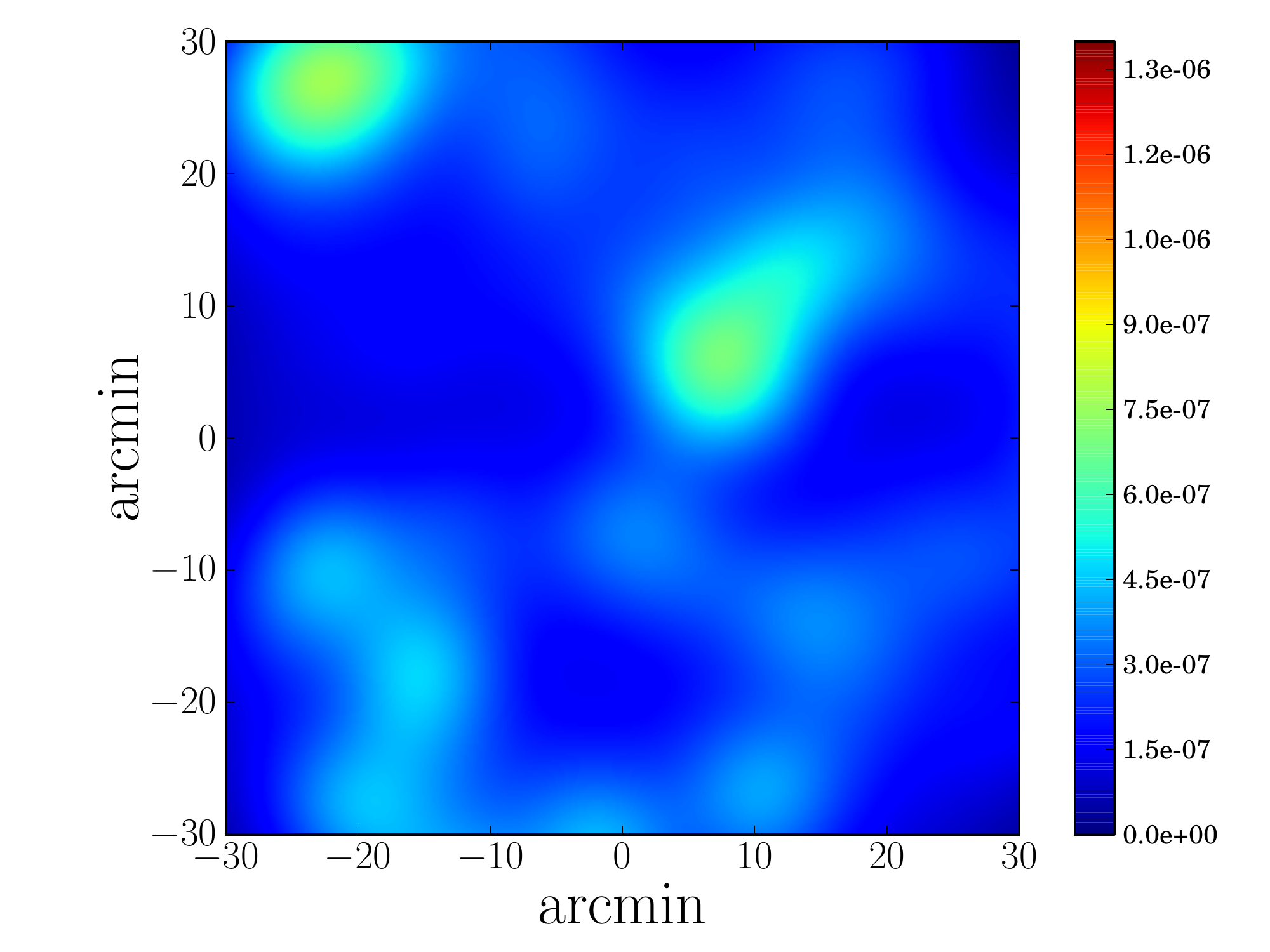}}
\resizebox{3.0in}{!}{\includegraphics[angle=0]{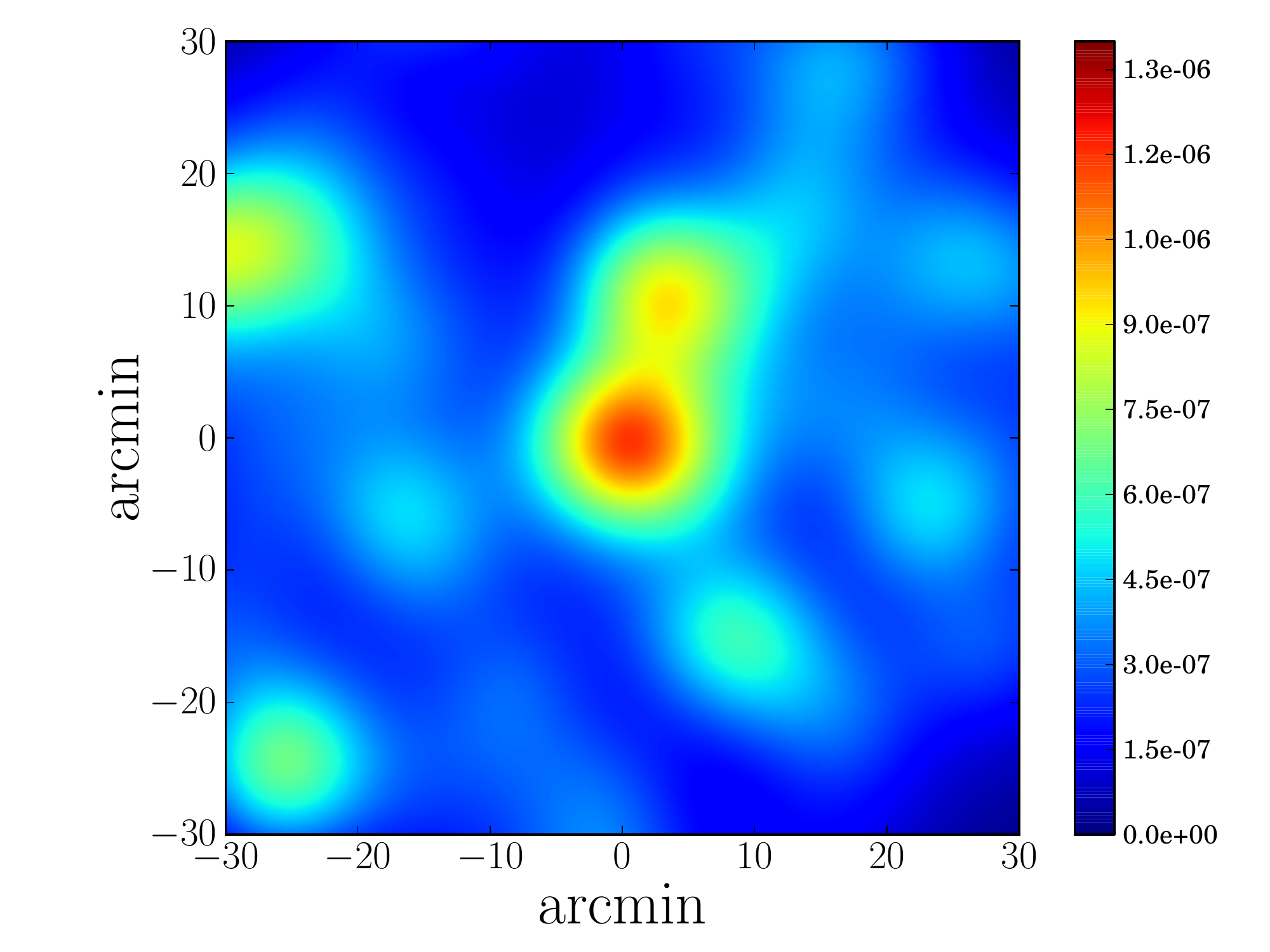}}
\resizebox{3.0in}{!}{\includegraphics[angle=0]{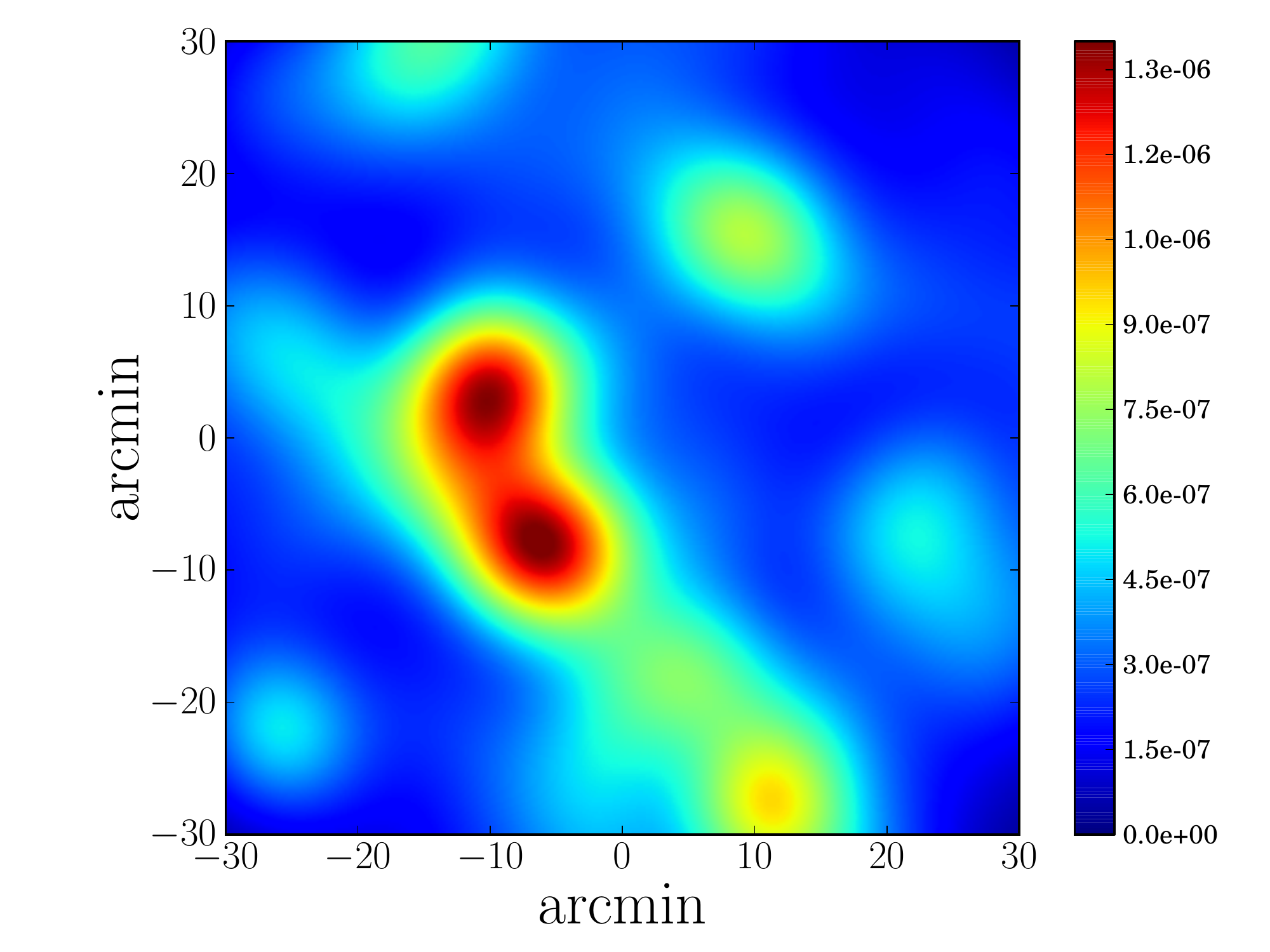}}
\resizebox{3.0in}{!}{\includegraphics[angle=0]{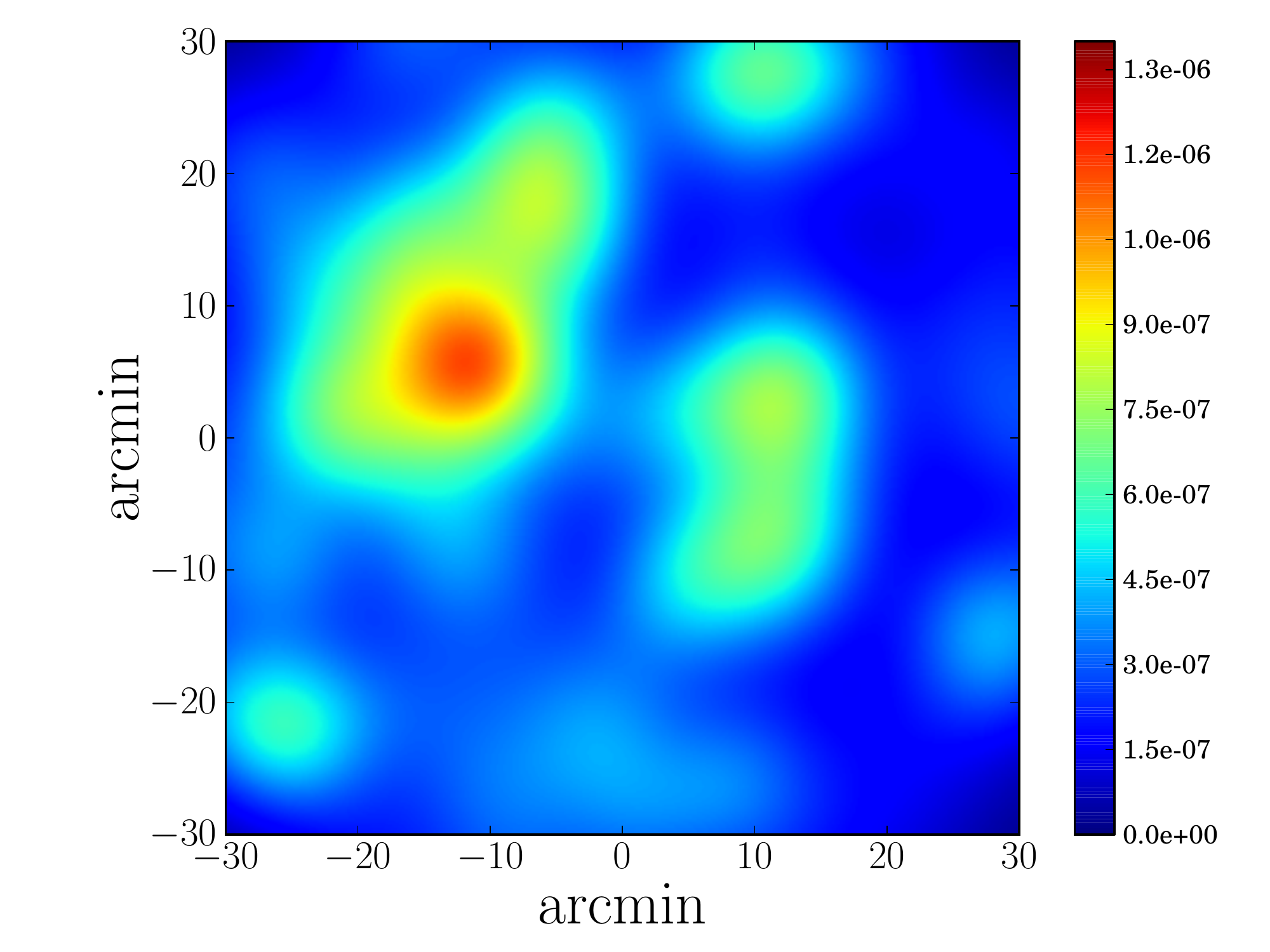}}
\resizebox{3.0in}{!}{\includegraphics[angle=0]{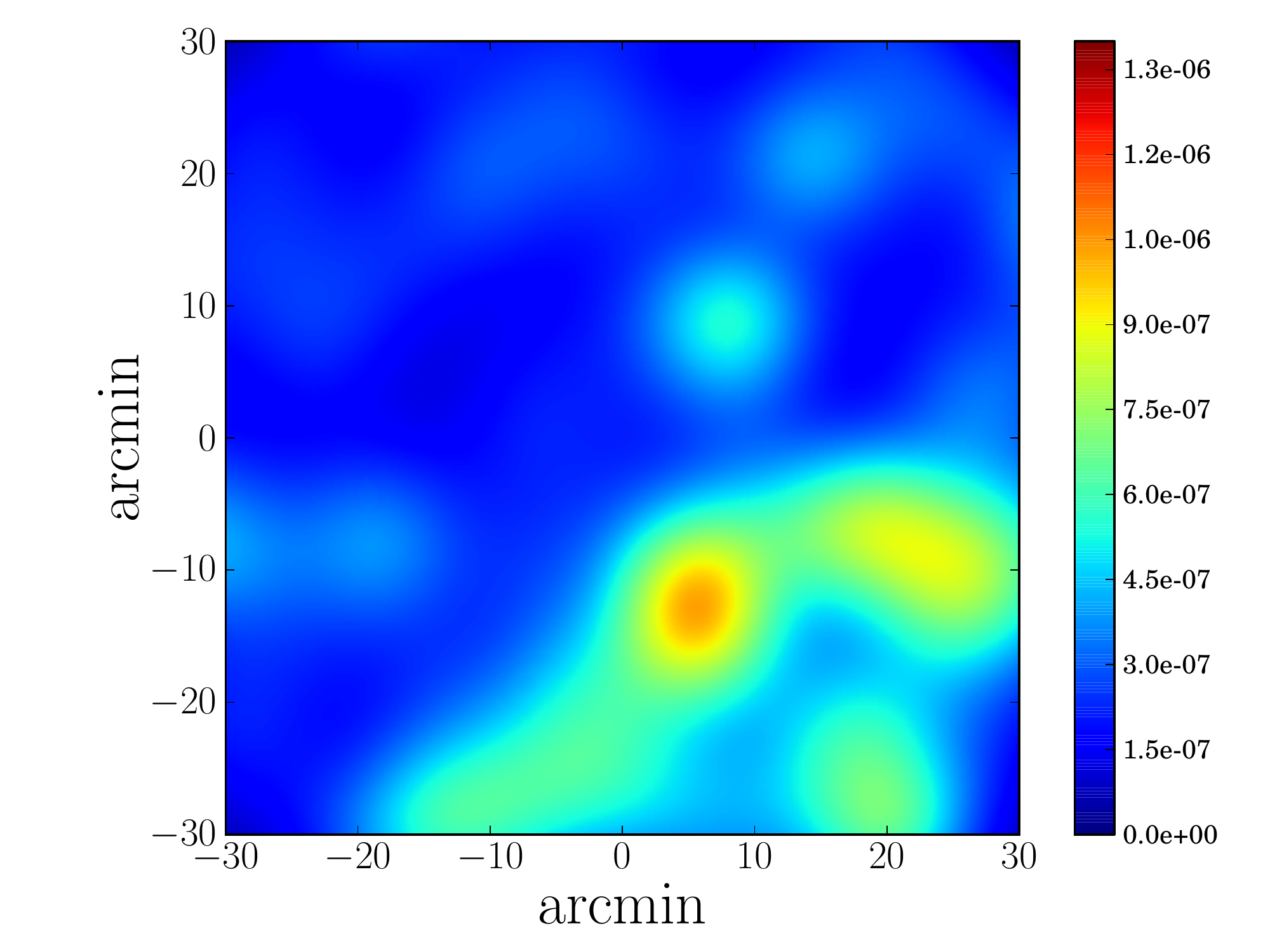}}
\vskip -0.0cm
\caption{
{\color{red}\bf Top-left panel} show the average Compton-y map of 10,000 individual maps centered on the quasar host
sampled from CS model at $z=1.4$. 
{\color{red}\bf The other five panels} show 
five randomly selected individual maps for five quasar halos. 
The pixel size is 0.034 arcmin.}
\label{fig:maps}
\end{figure}

We first validate the quasar models by comparing them to Planck observations.
Figure~\ref{fig:maps} shows Compton-y maps for five randomly selected quasar maps at $z=1.4$ (including the projection effects)
in the five panels other than the top-left panel and
the averaged over 10,000 such individual maps is shown in the top-left panel.
Each individual map is centered on the quasar halo from the CS model.
Halos that contribute to the signal are in the mass range we describe above and 
in some cases the quasar halo itself does not contribute to the signal if its mass fall outside the mass range.

\begin{figure}[!h]
\centering
\vskip -0.0cm
\hskip -0.4cm
\resizebox{2.35in}{!}{\includegraphics[angle=0]{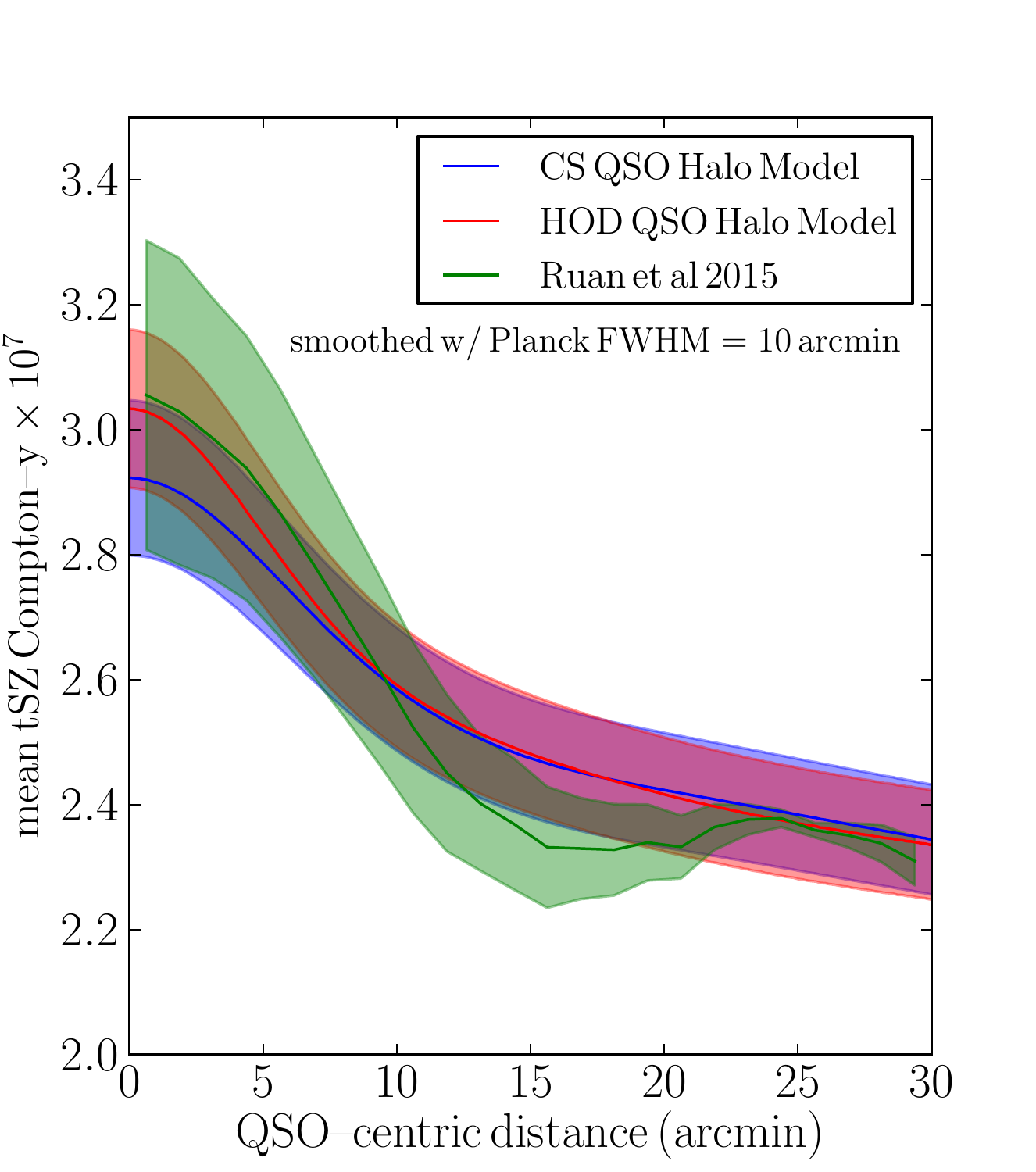}}
\hskip -0.6cm
\resizebox{2.35in}{!}{\includegraphics[angle=0]{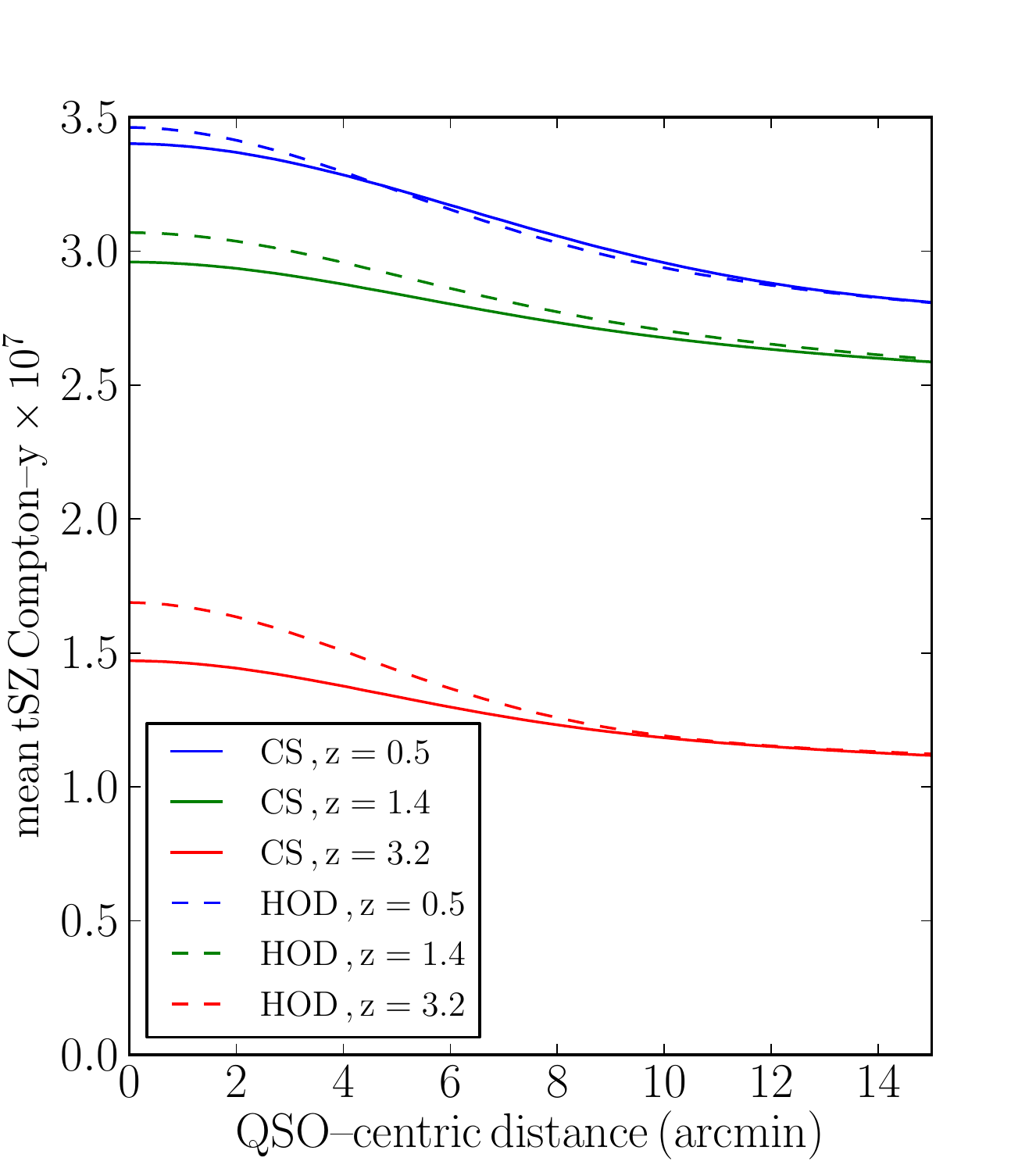}}
\hskip -0.65cm
\resizebox{2.35in}{!}{\includegraphics[angle=0]{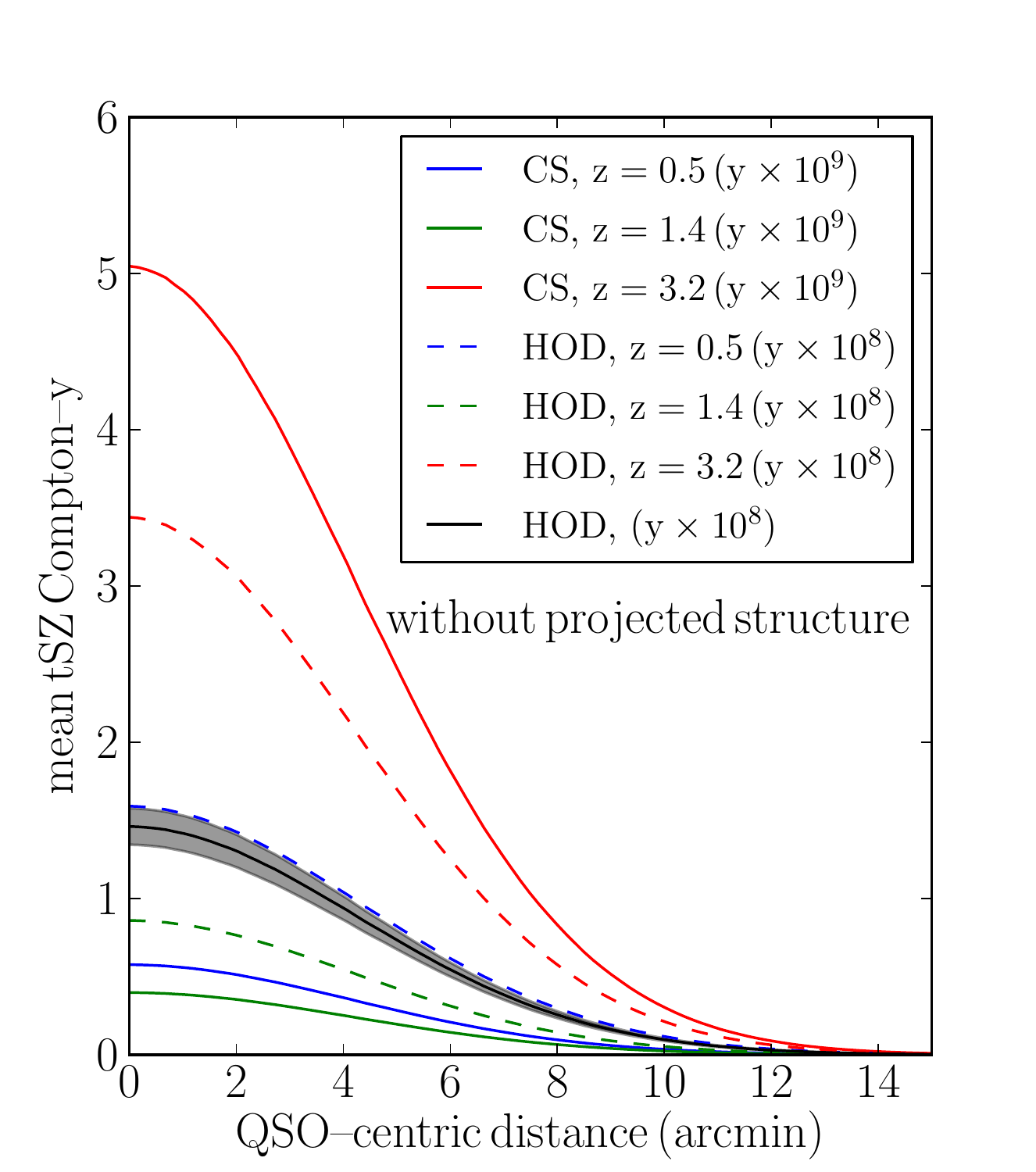}}
\vskip -0.0cm
\caption{
{\color{red}\bf Left panel} shows the mean Compton-y profile of quasars, including projection effects,
for the CS (blue shaded region) 
and HOD (purple shaded region) model, respectively, for a synthetic sample of quasars with redshift distribution 
[$z=(0.1,3.0)$ with median redshift of 1.5] mimicing that of the observed sample used in \citet[][]{2015Ruan}.
The radial profile of the observed quasars \citep[][]{2015Ruan}
is overplotted as the green shaded region without dust correction. 
We have normalized the observed radial profile with that of the models at 30 arcmin radius, 
which corresponds to the ``background".
{\color{red}\bf Middle panel} shows the mean Compton-y profile of a quasar, including projection effects,
at three separate redshifts, $z=0.5$ (blue curves), $z=1.4$ (green curves) and $z=3.2$ (red curves),
for the CS (solid curves) and HOD (dashed curves) model separately.
{\color{red}\bf Right panel} is similar to the middle panel, except that 
only the quasar-hosting halo contributes to the thermal energy in the map, 
without considering the contribution from other halos due to projection effects.
Also shown in black is the mean Compton-y profile in HOD model, without projection effects,
with appropriate weightings in accordance with 
that of the observed sample used in \citet[][]{2015Ruan}.
}
\label{fig:PLANCK}
\end{figure}

The left panel of 
Figure~\ref{fig:PLANCK} shows the Compton-y radial profile obtained by sampling for CS (blue shaded region) and HOD (purple shaded region)
model, respectively. 
Overplotted is the result obtained by stacking the Planck tSZ maps 
for quasars in the redshift range $(0.1,3.0)$ with median redshift of 1.5
\citep[green shaded region,][]{2015Ruan}.
To compare with Planck tSZ maps, we smooth our synthetic maps with a beam of FWHM=$10$~arcmin.
We see that, at the resolution of Planck of FWHM=$10$~arcmin, 
both CS and HOD model are consistent with the observed level of tSZ being contributed entirely
by shocked heated, virialized gas within massive halos.
Given our generous mass limit of contributing halos and neglect of gravitationally 
shock heated gas outside the virial radius, 
it is likely that the estimates for CS and HOD models
shown in the left panel of Figure~\ref{fig:PLANCK} are somewhat under-estimated.
Thus, in disagreement with \citet[][]{2015Ruan} with respect to feedback energy from other non-gravitational sources,
we see little evidence for a need of a large contribution to the tSZ 
from non-gravitational energy sources, including quasars or stars.

To better understand this discrepancy, we show in right panel of Figure~\ref{fig:PLANCK} 
the Compton-y profile in HOD model,
when only the quasar-hosting halo contributes to the thermal energy in the map,
neglecting the contribution from clustered neighboring halos.
We see that the isolated quasar map yields a tSZ signal peaked around
$y\sim 1.4\times 10^{-8}$ (black curve with shaded area)
versus $y\sim 3.0\times 10^{-7}$ as seen in the left panel where all neighboring halos are included.
It is hence very clear that the overall Compton-y parameter
reflects the collective thermal energy contribution of halos clustered around the quasar hosting halos in both CS and HOD models.
The collective effect exceeds that of the quasar host halo by more than an order of magnitude.
We attribute the suggested need of additional quasar feedback energy in order to account for the observed tSZ effect proposed by 
\citet[][]{2015Ruan} to the fact that projection effects due to clustered halos are not taken into account in their analysis.
In right panel of Figure~\ref{fig:PLANCK} we also show the 
mean tSZ signals for quasars at three different redshifts separately.
Since in this case no projected structures are included, the results are commensurate  
with the quasar halo masses in the models that increase with increasing redshift.
In the (HOD, CS) model \citep[][]{2015Cen},
the lower mass threshold of quasar hosts is 
$[2\times 10^{13},(2-5)\times 10^{12}]\msun$ at $z=3.2$,
$[5.8\times 10^{12},(2-5)\times 10^{11}]\msun$ at $z=1.4$
and 
$[5.7\times 10^{12},(1-3)\times 10^{11}]\msun$ at $z=0.5$.
It is also worth noting that, in the absence of projection effects,
the quasar tSZ signal in the CS model is about a factor of 5 (at $z=3.2$) 
to 25 (at $z=0.5$) lower than in the HOD model, due to differences in the quasar host halo masses
in the two models.

{\color{red}
It should be made clear that the projection effects are present at all redshifts.
In the middle panel of Figure~\ref{fig:PLANCK} 
we show the average Compton-y profile per quasar for the CS (solid curves) 
and HOD (dashed curves) model separately at $z=0.5$ (blue), $z=1.4$ (green) and $z=3.2$ (red),
including projection effects.
Two trends are seen and fully understandable.
First, overall, the tSZ signal per quasar, with projected structures,
increases with decreasing redshfit, in the range from $z=0.5$ to $z=3.2$.
This is expected due to continued growth of cosmic structure with time.
We note that, if we had not removed the most massive clusters in our tSZ maps
(to account for the masking-out of massive clusters in Planck maps \citep[][]{2014Planck},
the increase with decreasing redshift would be stronger.
Second, the ratio of tSZ signal with projection effects to that without projection effects
increases strongly with decreasing redshift, 
due to the combined effect of decreasing quasar host halo mass
and increasing clustering around massive halos with decreasing redshift.

In the left panel of Figure~\ref{fig:PLANCK} the observed y-map values are not dust-corrected.
The correction amplitude for dust effect with the procedure used by \citet[][]{2015Ruan}, 
by applying the channel weights from the \citet[][]{2014Hill} y-map construction 
to dust-like (modified blackbody) spectra, depends sensitively on dust temperature assumed.
Greco \& Hill (2015, private communications) show that
for a dust temperature of $34~$K used in \citet[][]{2015Ruan}, 
the y-map response is indeed negative over the entire redshift range of the quasar sample,
resulting in an increase in total thermal energy in the y-map by about $37\%$;
for lower dust temperatures, the y-map response becomes less negative and 
could go positive below some temperature for all redshifts;
for dust temperature of $20~$K, the y-map response is very slightly negative at $z<1.4$ but significantly positive
at $z>1.4$, with the net y-map response for the quasar sample slightly positive. 
With regard to dust temperature, observational evidence is varied but data suggesting lower temperatures are widespread. 
For example, \citet[][]{1998Schlegel} indicate dust temperature of $17-21~$K in our own Galaxy;
\citet[][]{2015Kashiwagi} suggest a dust temperature of $18~$K for dust around galaxies from far-infrared image stacking analysis;
\citet[][]{2014Greco} suggest an overall dust temperature of $20~$K in modeling the cosmic infrared background.
Thus, the contribution of dust emission itself to y-map depends significantly on the dust temperature
and the exact temperature of dust is uncertain at best and the actual y-map response is thus uncertain.
Even if we take the dust-corrected y-map from 
\citet[][]{2015Ruan}, given our results that the dust-uncorrected y values can
be explained soley by gravitational energy of halos hosting QSOs and neighboring ones, 
the QSO contribution is at most about $1/4$ of what is inferred in \citet[][]{2015Ruan}. 
}

\section{Testing Competing Quasar Models with Arc-Minute Resolution Thermal Sunyaev-Zeldovich Effect Maps}

\begin{figure}[!h]
\centering
\vskip -0.0cm
\hskip -1.0cm
\resizebox{3.5in}{!}{\includegraphics[angle=0]{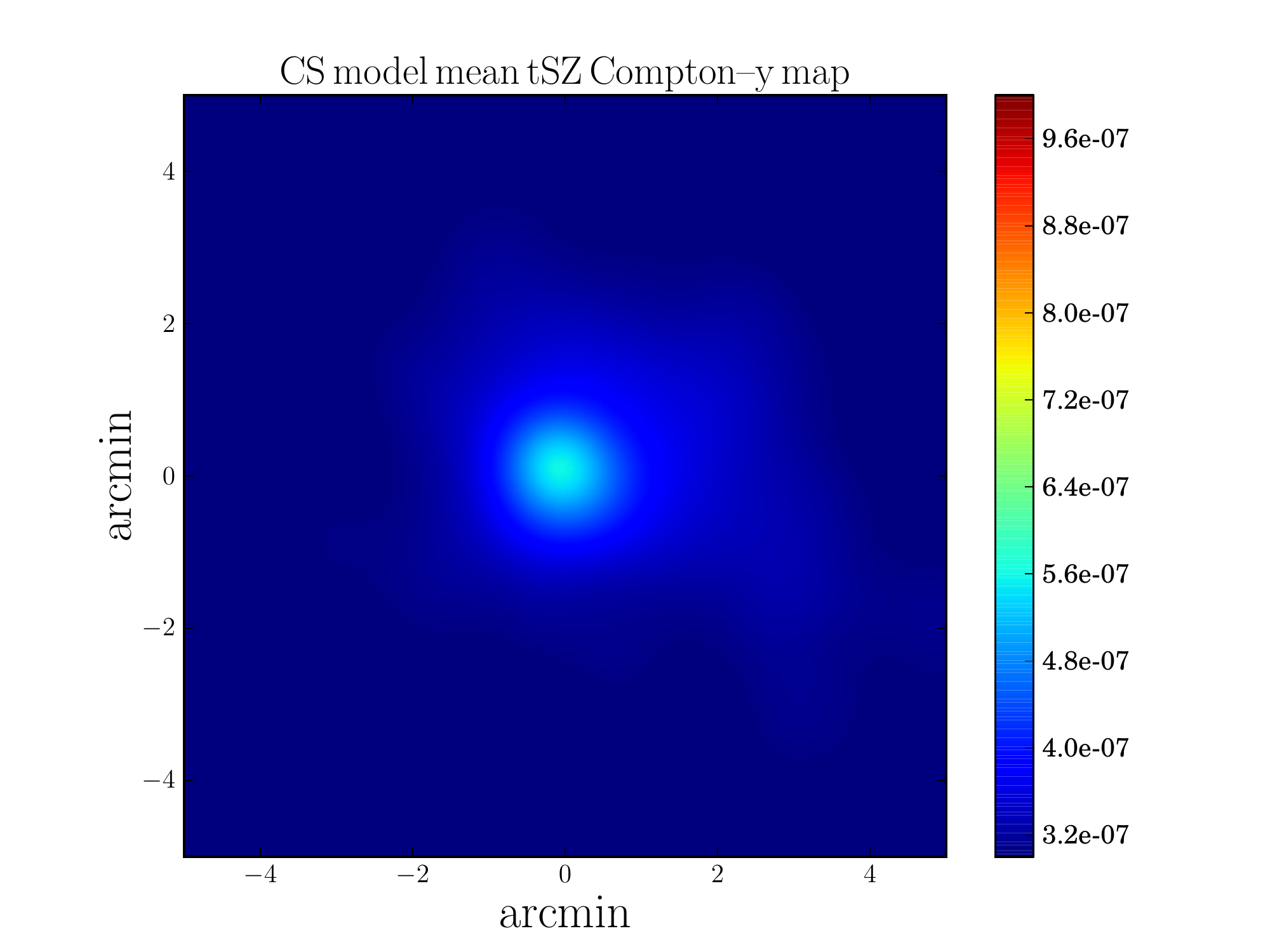}}
\hskip -1.0cm
\resizebox{3.5in}{!}{\includegraphics[angle=0]{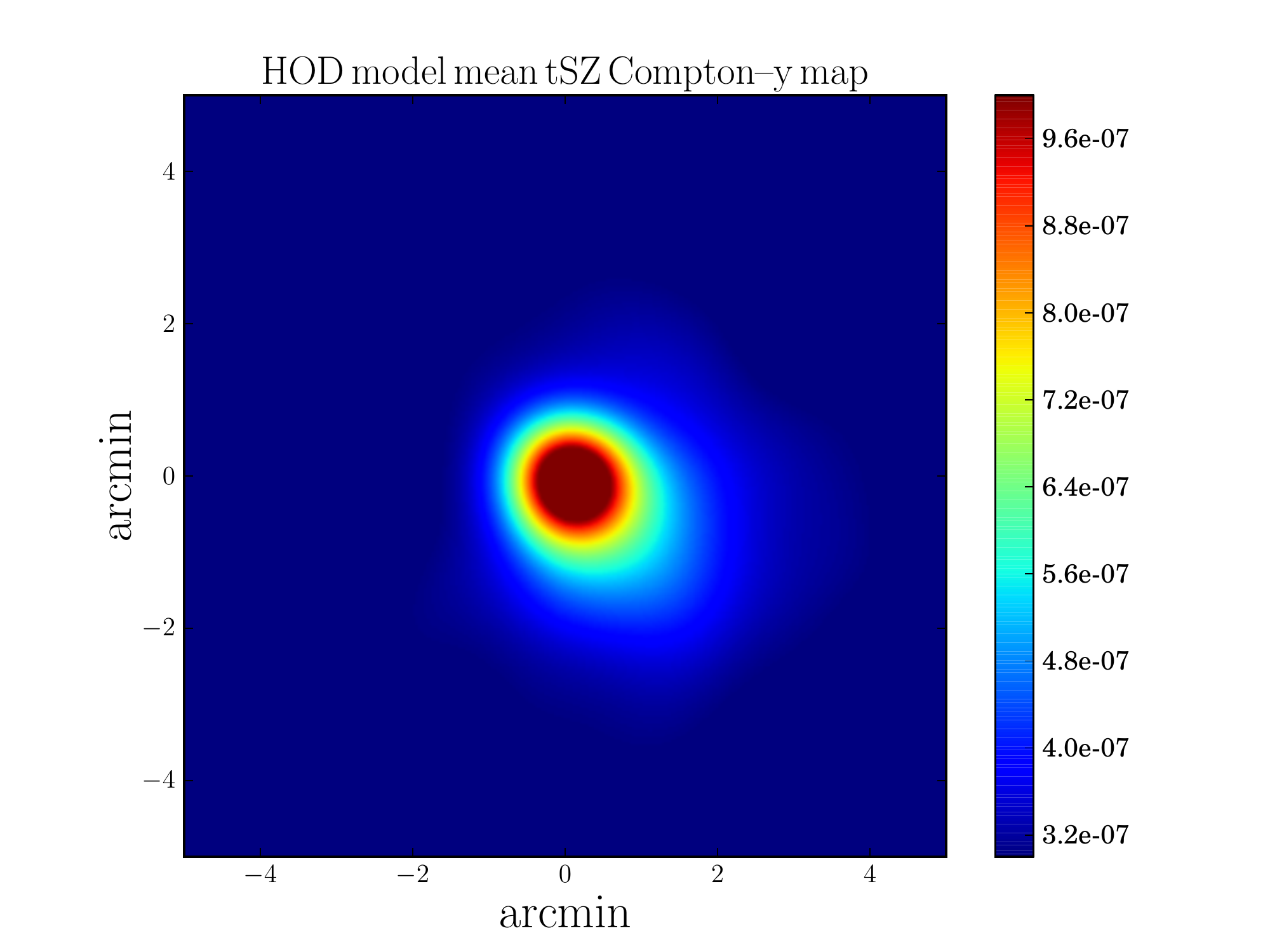}}
\vskip -0.0cm
\caption{
{\color{red}\bf Left panel} shows the stacked tSZ map of quasars with a median redshift of $1.5$ smoothed with FWHM=$1$~arcmin 
in the CS model. {\color{red}\bf Right panel} shows the same for HOD model.}
\label{fig:ACTmaps}
\end{figure}

Having validated both the CS and HOD models by the Planck tSZ data on 10~arcmin scales in the previous section,
here we propose a test to differentiate between them.
Figure~\ref{fig:ACTmaps} shows 
the stacked tSZ map of quasars with a median redshift of $1.5$ smoothed with FWHM=$1$~arcmin 
in the CS (left panel) and HOD (right panel) model.
The difference between the two model is visually striking:
the HOD model, being hosted by much more massive halos than the CS model,
displays a much more peaked tSZ profile at the arcmin scales.
The reason is that one arcmin corresponds to 516 kpc at $z=1.5$,
indicating that individual quasar hosting halos of mass $\ge 10^{13}\msun$ in the HOD model
are no longer significantly smoothed out by a $1$ arcmin beam.
The quasar hosting halos in the CS model, on the other hand,
have much lower masses than those in the HOD model
and hence have much lower tSZ effect at the arcmin scale.
At the arcmin scale, projection effects are much reduced compared to the $10$ arcmin scale.

\begin{figure}[!h]
\centering
\vskip -0.0cm
\resizebox{4.0in}{!}{\includegraphics[angle=0]{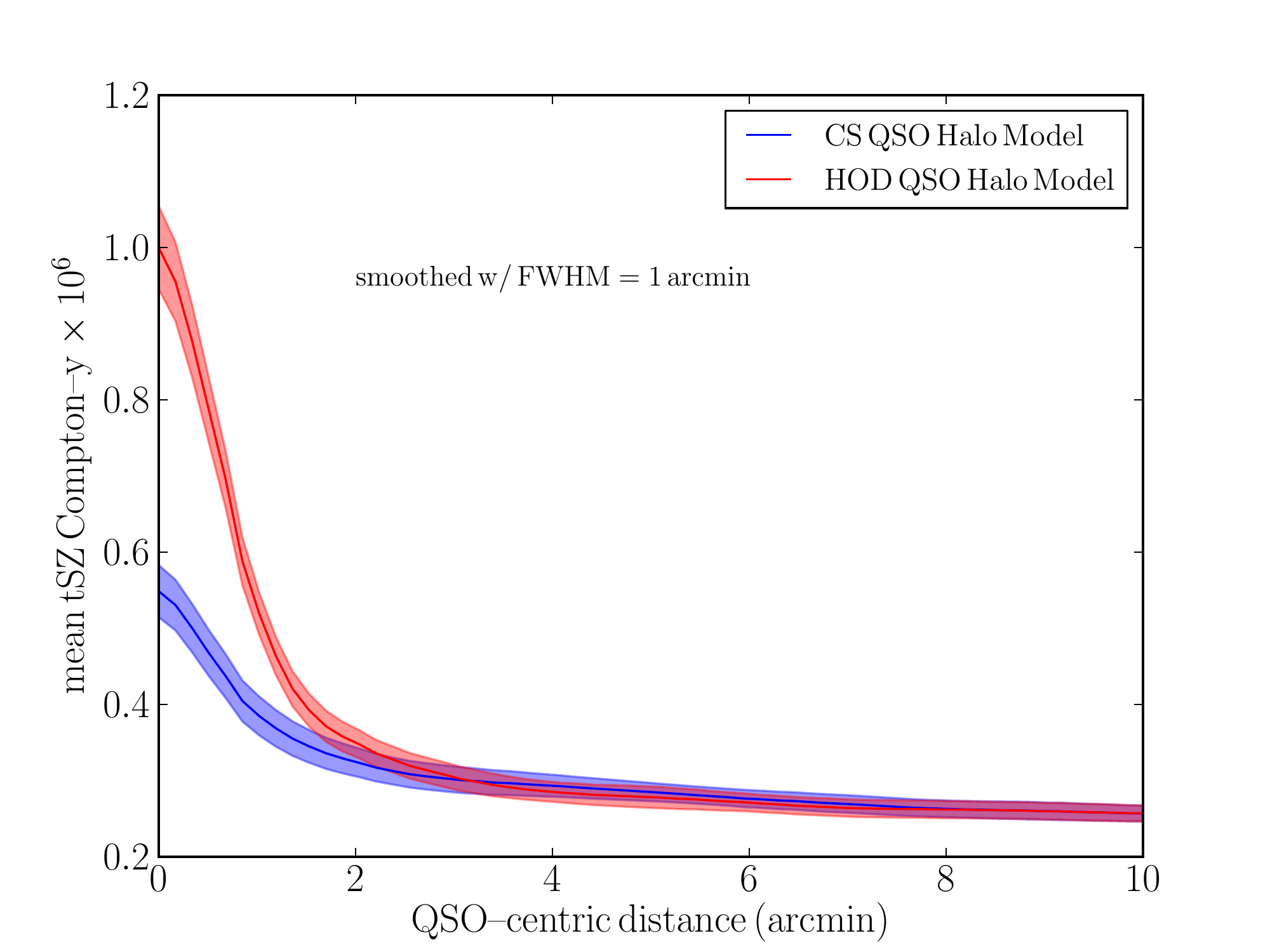}}
\vskip -0.0cm
\caption{
shows the predicted radial Compton-y profile of CS and HOD model smoothed with $1$ arcmin FWHM.}
\label{fig:ACT}
\end{figure}

Figure~\ref{fig:ACT} quantifies what is seen in Figure~\ref{fig:ACTmaps}
for the two quasar models.  
We see that, with FWHM=$1$ arcmin,
the central value of $y$ parameter differs by a factor of about two in the two models:
$(1.0\pm 0.05)\times 10^{-6}$ in the HOD model versus 
$(0.55\pm 0.03)\times 10^{-6}$ in the CS model.
This is a large difference and can be easily tested.

\begin{figure}[!h]
\centering
\vskip -0.0cm
\resizebox{4.5in}{!}{\includegraphics[angle=0]{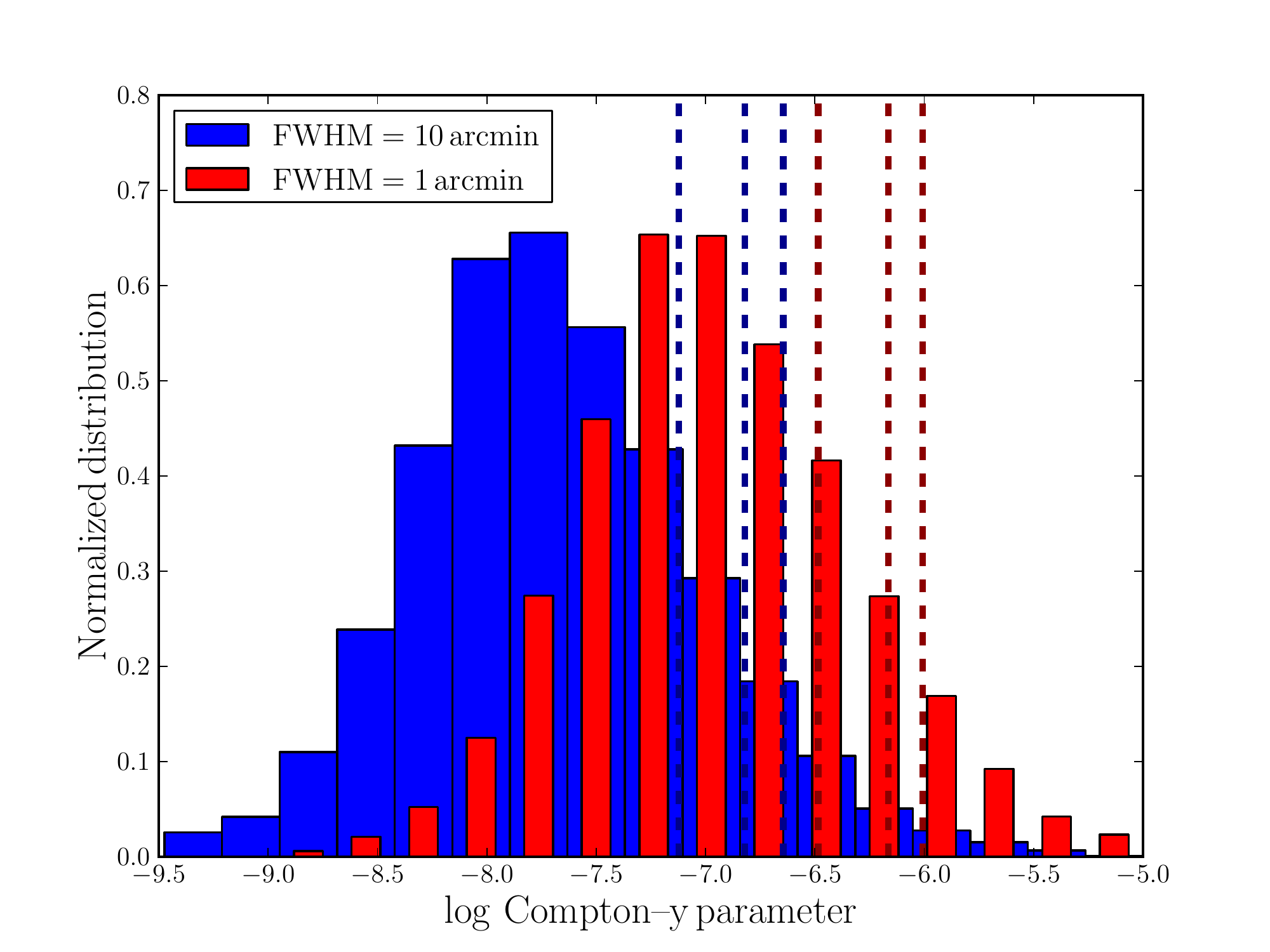}}
\vskip -0.0cm
\caption{
shows the probability distribution function (PDF) 
of $y$ parameter of the central region of radius $1$~arcmin 
of 10,000 individual quasar hosting halos (including projection effects) smoothed with FWHM=$1$ arcmin (red histogram) 
and smoothed with FWHM=$10$ arcmin (blue histogram) in the CS model.
The vertical lines color indicate the median and 
inter-quartile of the contribution to the mean y value of similar color histograms.}
\label{fig:hist}
\end{figure}

Before quantifying how the two models may be differentiated,
it is useful to understand the distribution of contributions from individual y maps to the averaged y map.
Figure~\ref{fig:hist} 
shows the probability distribution function (PDF) 
of y parameter of the central region of radius $1$~arcmin 
of 10,000 individual quasar hosting halos (including projection effects) smoothed with FWHM=$1$ arcmin (red histogram) 
and smoothed with FWHM=$10$ arcmin (blue histogram).
It is evident that the distribution of ${\rm \log y}$ in both cases is close to gaussian
hence the distribution of y is approximately lognormal. 
This indicates that the overall contribution to the stacked maps is skewed to the high end of the y distribution.
We find that 7.8\%, 12.1\% and 22.5\% of high $y$ quasar halos 
contribute to $25\%$, $50\%$ and $75\%$ of the overall $y$ value in the case with FWHM$=1~$arcmin,
6.3\%, 9.6\% and 18.4\% in the case with FWHM$=10~$arcmin.
Given the non-gaussian nature, we use bootstrap to estimate errors on the mean y value.
We find that the fractional error on the mean,  computed by 
bootstrap sampling from our 10,000 samples, 
is 3.7\% and 3.2\% for FWHM=10 arcmin and 1 arcmin cases, respectively.
Thus, with a sample of $26,000$ quasars as in \citet[][]{2015Ruan},
the fractional error on the mean would be 2\% for FWHM=1 arcmin case. 
Since the fractional difference between 
the HOD (${\rm y_{central}=(1.0\pm 0.05)\times 10^{-6}}$)  
and the CS (${\rm y_{central}=(0.55\pm 0.03)\times 10^{-6}}$ is 60\%,
this means that the HOD and CS model can be distinguished at $\sim 30\sigma$ level,
{\it if statistical uncertainties are the only uncertainties}.
It is thus likely that the significance level of differentiating the two models
using arcmin scale tSZ effect around quasars
will be limited by systematic uncertainties.

{\color{red}
As stated in \S 3,
there is a possibility that a significant fraction ($\sim 25\%$) of the 
observed thermal energy based on y-maps 
may be due to non-gravitational heating, such as quasar feedback suggested by \citet[][]{2015Ruan}.
Under the reasonable assumption that the energy from quasar feedback accumulates over time,
say via episodic high-energy radio jets,
the quasar feedback energy would be proportional to the galaxy stellar mass or approximately the halo mass,
given the observed correlation between supermassive black hole mass and the bulge stellar mass or velocity dispersion 
\citep[e.g.,][]{1998Magorrian,1998Richstone,2000Gebhardt,2000Ferrarese,2002Tremaine}.
If we further assume that the radial profile of the deposited energy from quasar feedback is the same as that of 
thermal energy sourced by gravitational energy,
it follows then that the central y-value of the (HOD,CS) model
would be boosted from [$(1.0\pm 0.05)\times 10^{-6}$, $(0.55\pm 0.03)\times 10^{-6}$]
shown in Figure~\ref{fig:ACT} to [$(1.4\pm 0.07)\times 10^{-6}$, $(0.77\pm 0.04)\times 10^{-6}$]. 
With the inclusion of this systematic uncertainty on quasar feedback energy,
the expected central y-value ranges would become 
[$(1.0-1.4)\times 10^{-6}$, $(0.55-0.77)\times 10^{-6}$], respectively,
for the (HOD,CS) model, which remain strongly testable with arcmin resolution tSZ observations.
}

\section{Conclusions}

We perform a statistical analysis of stacked $y$ maps of quasar hosts using 
{\em Millennium Simulation}.
Two significant findings may be summarized.
First, at the available resolution of FWHM=$10$ arcmin obtained by Planck data,
the observed tSZ effect can be entirely accounted for and explained by thermal energy of halos sourced by gravitational collapse.
No additional energy source is required at this conjunction.
It must be noted that at FWHM=$10$ arcmin projection effects are important with contribution to $y$ parameter by clustered halos 
with the $\sim 10$ arcmin scale dominating over the host halos themselves by an order of magnitude.
Considering uncertainties of dust temperature in the calibration of observed y-maps,
the maximum quasar feedback energy is about 25\% of that suggested \citep[][]{2015Ruan}.

Second, we show that, at FWHM=$1$ arcmin beam,
the central value of $y$ parameter 
is $(1.0\pm 0.05)\times 10^{-6}$ and 
$(0.55\pm 0.03)\times 10^{-6}$ in the HOD and CS model, respectively,
because of the significant differences in the masses of quasar hosting halos in the two models.
At $z\sim 0.5-2$, the host halos in the CS model have masses of $\sim 10^{11}-10^{12}\msun$,
compared to $(0.5-2)\times 10^{13}\msun$ in the HOD model.
With an observational sample of $26,000$ quasars, one will be able to distinguish between the HOD and CS models 
at a very high confidence level statistically ,
indicating that that the significance level will only be limited by systematic uncertainties.
With possible quasar feedback,
the expected central y-value uncertainty ranges would be enlarge to become 
[$(1.0-1.4)\times 10^{-6}$, $(0.55-0.77)\times 10^{-6}$], respectively,
for the (HOD,CS) model, which remain strongly testable with arcmin resolution tSZ observations.
Upcoming observations, such as Advanced ACT \citep[][]{2014Calabrese}, may be able to provide a definitive test.

\vskip 1cm

We are grateful to Dr. Ruan for sending us the data and useful discussion.
We also would like to thank Dr. Colin Hill for useful discussion.
This work is supported in part by grant NASA NNX11AI23G.
The Millennium simulation data bases used in this paper and the web application 
providing online access to them were constructed as part of the activities of the German Astrophysical Virtual Observatory.


\end{document}